\documentclass[english,twocolumn,superscriptaddress,longbibliography]{revtex4-1}

\usepackage{xcolor}
\usepackage{graphicx}
\usepackage{url}
\usepackage{amsmath}
\usepackage{amsfonts}
\usepackage{color}
\usepackage{babel}

\newcommand{\bs}[1]{\boldsymbol{#1}}

\begin{document}

\title{Design of double- and multi-bend achromat lattices with large
  dynamic aperture and approximate invariants}
 
\author{Yongjun Li}\thanks{email: yli@bnl.gov} \affiliation{Brookhaven
  National Laboratory, Upton 11973, New York, USA}

\author{Kilean Hwang}\thanks{currently at Michigan State Uni.}
\affiliation{Michigan State University, East Lansing, Michigan, 48824, USA}

\author{Chad Mitchell}\thanks{email:chadmitchell@lbl.gov}
\affiliation{Lawrence Berkeley National Laboratory, Berkeley 94720,
  California, USA}

\author{Robert Rainer} \affiliation{Brookhaven National Laboratory,
  Upton 11973, New York, USA}

\author{Robert Ryne} \affiliation{Lawrence Berkeley National Laboratory,
  Berkeley 94720, California, USA}
\author{Victor Smaluk} \affiliation{Brookhaven National Laboratory,
  Upton 11973, New York, USA}
 
\begin{abstract}
  A numerical method to design nonlinear double- and multi-bend achromat
  (DBA and MBA) lattices with approximate invariants of motion is
  investigated. The search for such nonlinear lattices is motivated by
  Fermilab's Integrable Optics Test Accelerator (IOTA), whose design is
  based on an integrable Hamiltonian system with two invariants of motion.
  While it may not be possible to design an achromatic lattice for a
  dedicated synchrotron light source storage ring with one or more exact
  invariants of motion, it is possible to tune the sextupoles and
  octupoles in existing DBA and MBA lattices to produce approximate
  invariants.  In our procedure, the lattice is tuned while minimizing the
  turn-by-turn fluctuations of the Courant-Snyder actions $J_x$ and $J_y$
  at several distinct amplitudes, while simultaneously minimizing
  diffusion of the on-energy betatron tunes. The resulting lattices share
  some important features with integrable ones, such as a large dynamic
  aperture, trajectories confined to invariant tori, robustness to
  resonances and errors, and a large amplitude-dependent
  tune-spread. Compared to the nominal NSLS-II lattice, the single- and
  multi-bunch instability thresholds are increased and the bunch-by-bunch
  feedback gain can be reduced.
\end{abstract}

\maketitle

\section{\label{sect:intro}introduction}

  The Integrable Optics Test Accelerator (IOTA)~\cite{antipov2017}, whose
  design is based on an integrable Hamiltonian system with two invariants
  of motion~\cite{danilov2008,danilov2010}, paves the way for a new class
  of highly nonlinear storage rings. Experiments using a lattice design
  with one invariant of motion have also been performed, both at IOTA and
  in the University of Maryland Electron Ring (UMER)~\cite{ruisard2019}.
  In each case, the lattice is tuned to provide one or more analytically
  known invariants of motion, resulting in a dynamic aperture (DA) that is
  large and robust to the presence of resonances.
  
  The storage rings used as dedicated synchrotron light sources are
  designed in a different way: a linear achromat lattice with a desired
  beam emittance is designed first, and then the nonlinear dynamics is
  optimized with sextupoles and/or octupoles. The nonlinear magnets are
  often tuned to control the low order resonance driving terms of the
  one-turn map~\cite{dragt2011} to obtain sufficient dynamic aperture.
  Under these conditions it is generally difficult, if not impossible, to
  optimize the nonlinear dynamics to produce a one-turn map with one or
  more exact invariants.  However, it is sometimes possible to produce
  approximate invariants, or quasi-invariants (QI), in these achromat
  lattices. This paper describes a procedure for designing near-integrable
  double-bend achromat (DBA) and multi-bend achromat (MBA) lattices with
  two QI. The motivation for constructing such lattices is that, although
  they are not completely integrable, the DA is large and robust to the
  presence of resonances. While crossing the resonance lines, their
  stop-band widths are observed to be narrow. Like nonlinear integrable
  lattices such as the one at IOTA, these lattices also have a large
  amplitude-dependent betatron tune-spread which can increase instability
  and space charge thresholds due to improved Landau
  damping~\cite{chao1993,stern2018}. As a result, the requirements on the
  feedback system's gain can be reduced significantly.  This research
  was motivated by related studies such as the square matrix
  method~\cite{yu2017} and the constant Courant-Snyder invariant
  method~\cite{borland2017,sun2017}.

  The remainder of this paper is outlined as follows:
  Section~\ref{sect:invar} explains the concept of Poisson-commuting
  invariants in integrable Hamiltonian systems, and describes a numerical
  approach for optimizing the nonlinear lattice to produce approximate
  invariants using symplectic tracking. Sections~\ref{sect:dba} and
  ~\ref{sect:mba} describe the properties of two such lattices, both of
  which have been constructed: the existing National Synchrotron Light
  Source-II (NSLS-II) DBA storage ring, and a diffraction-limited MBA
  storage ring (whose design is preliminary). Some detailed studies of the
  DBA lattice are described in Sect. \ref{sect:period}.  Section
  ~\ref{sect:decoh} describes the simulation of a kicked beam to
  illustrate the decoherence effect that results from a large nonlinear
  tune-spread.  Some discussion and a brief summary are given in
  Sect.~\ref{sect:summary}.  A technique to modify the action-like
  invariants to reshape the invariant tori (and the resulting DA) is
  described in the Appendix.

\section{\label{sect:invar}Lattice design procedure}
  
  A Hamiltonian system is {\it Liouville integrable} if it possesses a
  maximal set of independent Poisson commuting invariants of motion.  For
  a system described by a symplectic map on a phase space of dimension
  $2n$, this means that there exist $n$ functions $f_j$ $(j=1,\ldots,n)$
  on the phase space such that: i) each $f_j$ is invariant under the map,
  ii) the Poisson brackets satisfy $[f_i,f_j]=0$, and iii) the set of
  gradient vectors $\{\nabla f_j:j=1,\ldots,n\}$ is linearly independent
  ~\cite{liouville1855,arnold1989}.  The behavior of trajectories for a
  completely integrable system is well-known, i.e., all its trajectories
  are confined to tori with well-defined and stable tunes.
  
  By ignoring radiation and longitudinal acceleration, a charged
  particle's transverse motion in a storage ring is a 4-dimensional
  Hamiltonian system, described by a symplectic one-turn map
  $\mathcal{M}$. If we let the canonical coordinates of the system be
  denoted $\bs{z}=(x,p_x;y,p_y)$, a quantity $f(\bs{z})$ is an invariant
  of the map $\mathcal{M}$ if:
  \begin{equation}
    f(\mathcal{M}({\bf z}))=f({\bf z}).
  \end{equation}
  If two such invariants $f_i$, $(i=1,2)$ exist, if they are independent:
  \begin{equation}
   \nabla f_1\times\nabla f_2\neq 0,
  \end{equation}
  and if they Poisson-commute:
  \begin{align}
   \left[f_1,f_2\right]&=
   \left(\frac{\partial f_1}{\partial x}\frac{\partial f_2}{\partial p_x}-
   \frac{\partial f_1}{\partial p_x}\frac{\partial f_2}{\partial x}\right) \notag \\
   &+\left(\frac{\partial f_1}{\partial y}\frac{\partial f_2}{\partial p_y}-
   \frac{\partial f_1}{\partial p_y}\frac{\partial f_2}{\partial y}\right)=0,
  \end{align}
  the lattice is Liouville integrable.
  
  When the map $\mathcal{M}$ is linear and uncoupled, the Courant-Snyder
  actions $J_x$ and $J_y$ form the most commonly-used Poisson commuting
  pair of invariants, where:
  \begin{align}\label{eq:quadraticJ}
    J_x = \frac{1}{2}(\bar{x}^2+ \bar{p}_x^2) =
    \frac{1}{2}\left(\gamma_xx^2+2\alpha_xxp_x+\beta_xp_x^2\right),
  \end{align}
  in the horizontal plane, with a similar expression for $J_y$.  Here
  $\alpha_x,\;\beta_x,\text{and}\;\gamma_x$ are the horizontal Twiss
  parameters~\cite{courant1958} at the longitudinal location where the
  Poincar\'{e} section is observed.  The canonical action-angle
  coordinates are $(\Phi_x,J_x,\Phi_y,J_y)$, where $\Phi_{x,y}$ denotes
  the betatron phase in each plane, and the one-turn map is determined by
  the phase advance completed in a single revolution:
  \begin{align}\label{eq:phaseAdvance}
    \phi_x&=\Phi_{x,i+1}-\Phi_{x,i}\nonumber\\ &=-
    \arctan\left({\frac{\bar{p}_{x,i+1}}{\bar{x}_{i+1}}}\right)+
    \arctan\left({\frac{\bar{p}_{x,i}}{\bar{x}_i}}\right)+k\cdot2\pi,
  \end{align}
  with a similar expression for $\phi_y$.  Here, $k$ is the integer part
  of the betatron tune.  The phase advance values $\phi_x$, $\phi_y$ are
  independent of the actions $J_x$ and $J_y$.

\begin{figure}[!ht]
    \centering \includegraphics[width=0.6\columnwidth]{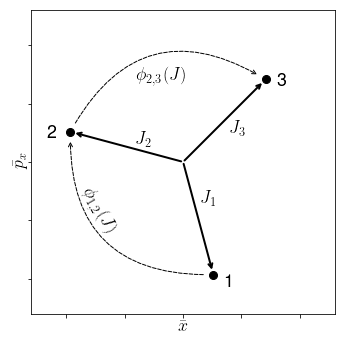}
    \caption{\label{fig:invariant} Schematic illustration of a rotating
      trajectory observed at a Pioncar\'{e} section with normalized
      coordinates ($\bar{x},\;\bar{p}_x$). The fluctuations of the action
      $J_{x}$ and phase advance $\phi_{x}$ observed in multi-turn tracking
      simulations are the objectives to be minimized.  A similar picture
      applied in the vertical plane.}
  \end{figure}

  In a realistic storage ring, once the linear lattice and the nonlinear
  magnet locations are fixed, the one-turn map $\mathcal{M}$ depends on
  the nonlinear magnet strengths $K_i,\;\text{with}\;i\geq2$.  It is
  difficult, if not impossible, to tune the $K_i$ so that $\mathcal{M}$
  possesses even one exact invariant.  However, we can imitate the linear
  case by constructing a nonlinear system in which the Courant-Snyder
  actions $J_x$, $J_y$ form a pair of approximate invariants, as
  illustrated in Fig.~\ref{fig:invariant}.  Unlike the linear case,
  however, the phase advance values $\phi_x$ and $\phi_y$ can depend on
  the actions $J_x$ and $J_y$.
  
\begin{figure}[!ht]
    \centering \includegraphics[width=0.7\columnwidth]{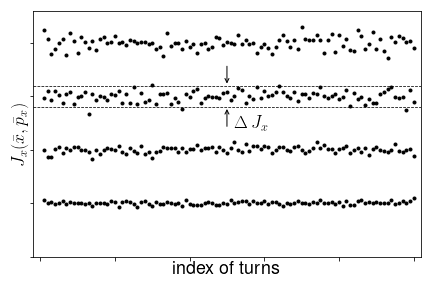}
    \caption{\label{fig:I1} Schematic illustration of the fluctuation of
      actions $\Delta J_x$ starting from different initial
      amplitudes. Usually the fluctuations increase gradually with the
      initial amplitude.}
  \end{figure}

  The procedure is as follows.  To optimize the behavior of the
  Courant-Snyder action $J_x$ within the available DA, multiple particles
  with different values of $J_{x,0}$ are launched. Element-by-element
  tracking of this set of particles is used to compute the turn-by-turn
  evolution of $J_x$.  The tracking is implemented with a kick-drift
  symplectic integrator~\cite{yoshida1990} to preserve the geometry of the
  Hamiltonian system. The available nonlinear knobs are simultaneously
  tuned to minimize the turn-by-turn fluctuations of $J_x$ for each
  particle, as illustrated in Fig.~\ref{fig:I1}.

\begin{figure}[!ht]
    \centering \includegraphics[width=0.7\columnwidth]{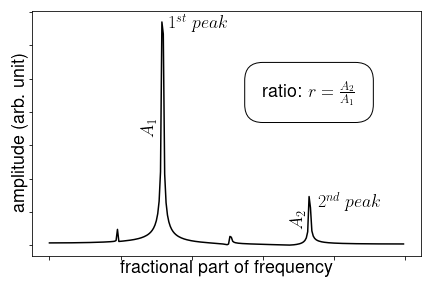}
    \caption{\label{fig:I2} Schematic illustration of the spectrum
      obtained from turn-by-turn trajectory data $\bar{x}\pm i\bar{p}_x$.
      The ratio between the amplitudes of the two leading frequencies
      $\frac{A_2}{A_1}$ is the objective to be minimized to suppress the
      orbit tune diffusion.}
  \end{figure}

  At the same time, we minimize the turn-to-turn variations of the
  horizontal phase advance. Instead of directly calculating the phase
  advance for each turn with Eq.\eqref{eq:phaseAdvance}, the complex
  turn-to-turn evolution of $\bar{x}\pm i\bar{p}_x$~\cite{chao2002} was
  analyzed in the frequency domain. One reason for using such a spectral
  method is to determine whether the fractional tune is below or above the
  half integer. The amplitudes of the two leading frequencies were
  computed utilizing the Numerical Analysis of Fundamental Frequencies
  (NAFF) technique~\cite{laskar2003}. By tuning the nonlinear knobs, the
  ratio between the two leading frequencies $r=\frac{A_2}{A_1}$ was
  minimized. As a consequence, the smaller amplitude frequencies were also
  suppressed (Fig.~\ref{fig:I2}). In principle, the summation of all
  non-fundamental peaks should be compared against the fundamental
  frequency peak, but the computational cost will become high. Since
  NAFF outputs the peaks in descending order, we only use the two leading
  peaks to get a quick but rough approximation. This procedure is
  performed independently for several initial conditions of varying
  amplitude.  As a result, the tune diffusion of each particle is
  suppressed, but the tunes may be amplitude-dependent.  The same
  procedure is repeated for the vertical plane.

  Since the goal is to minimize the fluctuations of four different
  quantities for different initial conditions simultaneously, the
  construction of such a nonlinear lattice becomes a typical
  multi-objective optimization problem:
  \begin{itemize}
    \item given a set of nonlinear knobs $K_i$ within their
      allowed ranges;
    \item subject to some constraints, such as maintaining certain desired
      chromaticities;
    \item simultaneously minimize the objective functions, i.e.,
      $\frac{\Delta J_{x,y}}{J_{x,y}}$ and
      $r_{x,y}=\frac{A_{2;x,y}}{A_{1;x,y}}$ of multi-particles launched
      from different initial conditions.
  \end{itemize}
  Multi-objective optimization techniques are now widely used in the
  accelerator community.  Here, the non-dominated sorting genetic
  algorithm (NSGA-II)~\cite{deb2001} was used. Five virtual particles with
  gradually increasing initial values of $J_{0;x,y}$ were launched, and
  for each initial condition, four objectives were used.  The total number
  of objectives was therefore $5\times 4=20$.

  Thus far, we have only discussed uncoupled linear lattices. When linear
  coupling is present, a different parameterization, such as the one
  described in \cite{edwards1973}, is needed.

\section{\label{sect:dba}Applied to double-bend achromat}

\begin{figure}[!ht]
    \centering \includegraphics[width=0.7\columnwidth]{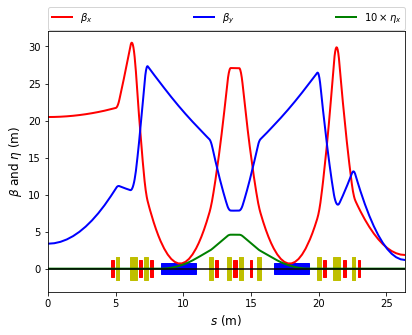}
    \caption{\label{fig:dbacell} Linear optics and magnet layout for
      one cell of the NSLS-II storage ring. The red blocks represent
      sextupoles. The strengths of six harmonic sextupoles were tuned
      during optimization of the nonlinear optics
      (Sect.~\ref{sect:invar}).}
    \end{figure}

  In this Section, we introduce a nonlinear DBA lattice for the NSLS-II
  main storage ring~\cite{bnl2013}, which is presently in operation at
  Brookhaven National Laboratory. It is a $3^{\text{rd}}$ generation
  medium energy (3 GeV) light source. The storage ring's lattice is a
  typical DBA structure with its main parameters listed in
  Table~\ref{tab:dba}. Its linear optics for one cell is illustrated in
  Fig.~\ref{fig:dbacell}. The whole ring is composed of 30 such cells. In
  this configuration, three families of chromatic sextupoles are used to
  correct its chromaticity to $+7$. Then, six families of harmonic
  sextupoles in dispersion-free sections are used as tuning knobs for the
  multi-objective optimization described in Section \ref{sect:invar}.

  \begin{table}[!hbt]
   \centering
   \caption{Main parameters of NSLS-II storage ring}
   \begin{tabular}{|p{0.4\columnwidth}|p{0.4\columnwidth}|}
     \hline
       \textbf{Parameters}       & \textbf{Values}  \\
       \hline
       \hline
       Hor. emit. ($nm$)         & 2.1              \\
       Natural chrom. (x/y)      & -101/-40         \\ 
       Tune (x/y)                & 33.22/16.26      \\ 
       Energy spread             & $5.1\times10^{-4}$\\
       Damp. partition (x/y/s)   & 1.0/1.0/2.0      \\
     \hline
   \end{tabular}
   \label{tab:dba}
  \end{table}

\begin{figure}[!ht]
     \includegraphics[width=0.48\columnwidth]{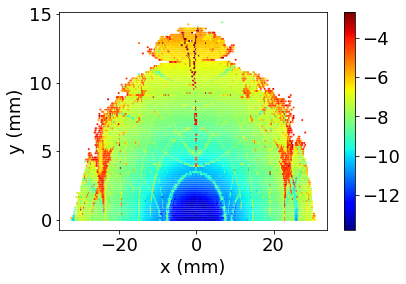}
     \includegraphics[width=0.48\columnwidth]{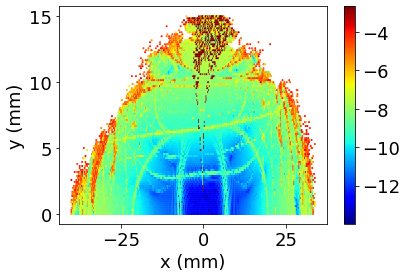}
     \caption{\label{fig:dbada} DA of the DBA lattice with QI (left) and the 
        NSLS-II lattice at nominal operation (right).
       The lattices have chromaticities of $\xi=7$ and $\xi=2$, respectively.
       Colors indicate the tune diffusion obtained with the NAFF
       technique.  }
  \end{figure}

  Below, we present the nonlinear lattice performance of an optimized
  solution using the tracking simulation code
  \textsc{elegant}~\cite{borland2000}. All the tracking simulations in
  this paper were performed with this code unless stated
  otherwise. Fig.~\ref{fig:dbada} illustrates on-momentum dynamic aperture
  (DA through 1,024 turns of particle tracking) observed at the center of
  the long straight section, i.e., $s=0$ in Fig.~\ref{fig:dbacell}. Each
  stable initial condition is colored with its tune diffusion
  $\log_{10}(\Delta\nu_x^2+\Delta\nu_y^2)$~\cite{laskar2003} obtained from
  turn-by-turn data using the NAFF algorithm. The nominal operation
  lattice's DA at chromaticity $\xi=2$ is also shown for
  comparison. Although stronger sextupoles are needed to correct the
  chromaticity in the QI lattice, the obtained DA is comparable with the
  nominal operation lattice.

\begin{figure}[!ht]
     \includegraphics[width=0.48\columnwidth]{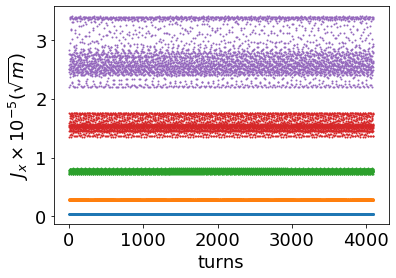}
     \includegraphics[width=0.48\columnwidth]{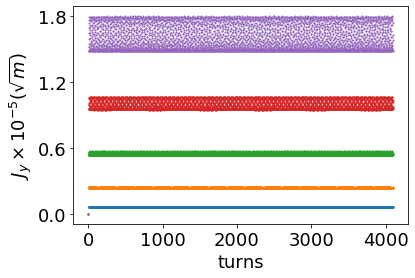}
     \caption{\label{fig:dbaI} Evolution of $J_{x,y}$ in the DBA
       lattice starting from 5 different initial conditions in the the
       horizontal (left) and vertical (right) planes.}
  \end{figure}

\begin{figure}[!ht]
     \includegraphics[width=0.48\columnwidth]{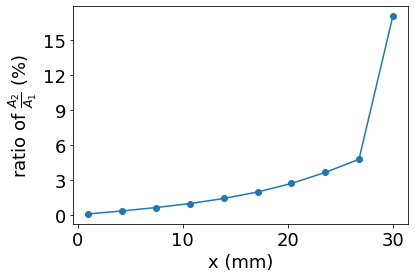}
     \includegraphics[width=0.48\columnwidth]{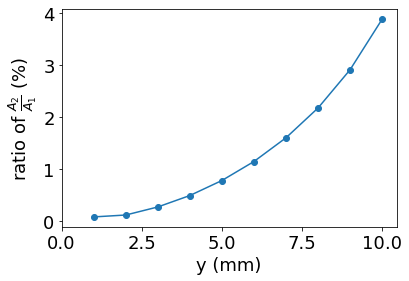}
     \caption{\label{fig:dbaratio} Ratio of the two leading NAFF
       components increases with the initial particle amplitude in the
       horizontal (left) and vertical (right) planes for the DBA lattice.}
  \end{figure}

  The turn-by-turn evolution of the Courant-Snyder actions $J_{x,y}$ for
  particles at 5 distinct amplitudes are shown in Fig.~\ref{fig:dbaI}. The
  size of the visible fluctuations increases gradually with the amplitude
  of the initial condition. The spectral analysis of tracking data using
  the NAFF technique indicates that some non-dominant frequencies
  gradually become stronger as well. The amplitude ratio between the two
  leading frequencies increases as shown in Fig.~\ref{fig:dbaratio}.

\begin{figure}[!ht]
     \includegraphics[width=0.48\columnwidth]{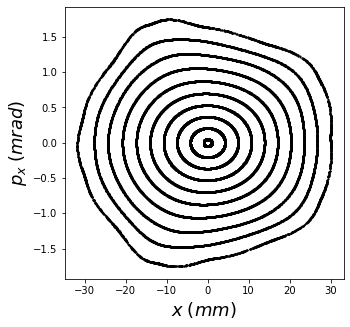}
     \includegraphics[width=0.48\columnwidth]{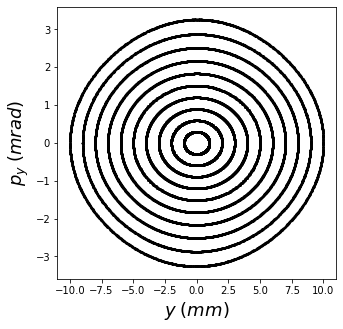}
     \caption{\label{fig:dbaphase} Simulated trajectories of the DBA
       lattice starting from different initial conditions in the
       horizontal (left) and vertical (right) phase space. Within the DA,
       although the trajectories deviate from the Courant-Snyder ellipse,
       they are still confined to thin tori.}
  \end{figure}

  One of the features of an integrable system is that the trajectories are
  confined to tori in the phase space.  This is apparent in the
  turn-by-turn tracking data shown in Fig.~\ref{fig:dbaphase}. Although
  trajectories begin to gradually deviate from the Courant-Snyder ellipse
  when the amplitude increases, they are still confined to deformed tori.
  It therefore appears that this lattice possesses two QIs whose values
  near the reference orbit are quantitatively close to the Courant-Snyder
  actions. This feature can also be observed from the symmetry of DA in
  the horizontal plane as shown in Fig.~\ref{fig:dbada}.
  
\begin{figure}[!ht]
     \includegraphics[width=0.48\columnwidth]{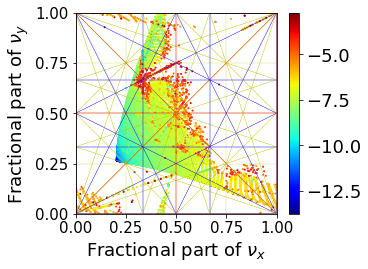}
     \includegraphics[width=0.48\columnwidth]{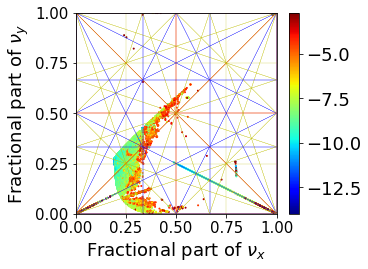}
     \caption{\label{fig:dbafma} Tune footprint of the NSLS-II DBA lattice
       with QI (left) and the nominal lattice for routine operation
       (right). The design working point is 33.22/16.26. Compared to the
       nominal lattice, a very large amplitude-dependent tune-spread is
       observed in the QI lattice, and various resonance lines can be
       crossed with narrow stop-band widths.  Here colors indicate the
       tune diffusion, as in Fig.~\ref{fig:dbada}.}
  \end{figure}

  Like the IOTA ring, this lattice also provides a large
  amplitude-dependent tune-spread as illustrated in the left subplot of
  Fig.~\ref{fig:dbafma}. This property can increase instability and space
  charge thresholds through improved Landau damping. Even while crossing
  the low order resonance lines such as $\nu_x=1/3$, the stop-band widths
  are observed to be much narrower than those in conventional nonlinear
  lattices~\cite{schoch1957, wiedemann2015}. The nominal operation
  lattice's tune footprint (as shown in the right subplot) can provide a
  smaller tune spread within the bunch when an instability occurs.
  
\begin{figure}[!ht]
     \includegraphics[width=0.48\columnwidth]{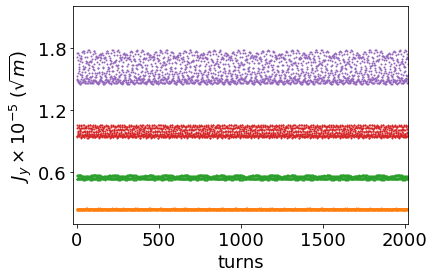}
     \includegraphics[width=0.48\columnwidth]{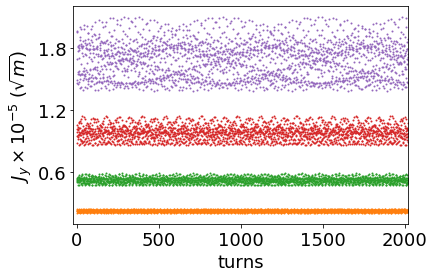}
     \caption{\label{fig:y_comp} Comparison of the vertical action
       fluctuations in the QI lattice (left) and the nominal NSLS-II
       lattice (right). Starting from the same initial conditions
       $(x,p_x,y,p_y)$, the action in the QI lattice has smaller
       fluctuations, despite the fact that the lattice has a high
       chromaticity $\xi=7$. The same feature can also be observed for the
       horizontal plane.}
  \end{figure}

  It is interesting to directly compare the fluctuations in actions for
  the QI lattice and the nominal lattice. Although their DAs are
  comparable, the fluctuations $\Delta J$ in the QI lattice are better
  suppressed as illustrated in Fig.~\ref{fig:y_comp}. When a small
  physical aperture exists in a ring, such as an undulator's gap, a
  trajectory with large action variation might be blocked. Then its
  realistic DA will be smaller than the physical aperture. In this case, a
  less fluctuating trajectory will be preferable.
  
\begin{figure}[!ht]
     \includegraphics[width=0.7\columnwidth]{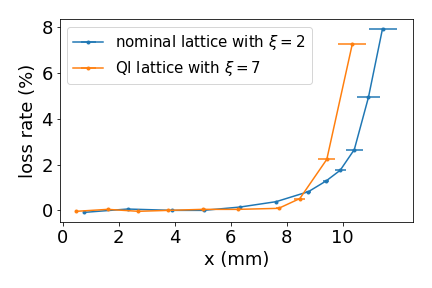}
     \caption{\label{fig:dbadaerr} Measured dynamic aperture of the DBA
       lattice with QI (yellow) and the NSLS-II lattice at nominal
       operation (blue).  The lattice chromaticities are $\xi=7$ and $\xi=2$,
        respectively.}
  \end{figure}

  The robustness of this lattice has been confirmed with a beam
  test. After loading its sextupoles settings, a near 100\% off-axis
  injection efficiency was achieved, which indicates its DA is sufficient
  for routine operation. Then, the horizontal DA was scanned by kicking
  the beam transversely to observe its loss rate as shown in
  Fig.~\ref{fig:dbadaerr}.  Although the measured DA is about
  $10\pm1\;mm$, which is worse than the simulation prediction due to
  various realistic errors, it still can satisfy the requirement
  ($>8.5\;mm$) for off-axis injection. The nominal operation lattice's DA
  is also shown for comparison.
  
  A bunch-by-bunch feedback (BBFB) system is frequently used in light
  source rings to suppress instabilities in the beam centroid motion.
  Limiting the BBFB gain is critical to avoid a degradation of the beam
  brightness.  For example, a significant vertical emittance increase was
  observed when an excess gain was applied~\cite{cheng2018} at the
  NSLS-II ring.  Although the beam centroid instabilities could be well
  suppressed with the BBFB system, the emittance blowup effect due to a
  large feedback gain was pronounced, especially at higher chromaticity.
  
  In this study, after accumulating a beam current of 400 mA, the BBFB
  system was re-optimized to reduce its gain gradually. Each bunch's
  turn-by-turn data was used for a spectral analysis to identify the
  appearance of sidebands of betatron motion caused by various
  instabilities. Compared with the NSLS-II lattice at nominal operation,
  the required gain for the nonlinear QI lattice was reduced by 50\% and
  75\% in the horizontal and vertical planes, respectively. This appears
  to be due to the increase in nonlinear tune-spread and chromaticity.

  The single bunch instability threshold was also increased in the QI
  lattice. For the nominal lattice, the charge threshold is around 2-3 mA
  without using BBFB, and 5-6 mA with BBFB~\cite{cheng2016}. For the QI
  lattice, this threshold was observed to be greater than 8 mA without
  BBFB, and 12 mA with BBFB. The actual threshold might be higher, because
  the single bunch accumulation was interrupted by high outgassing due to
  the extensive single bunch radiation pulse.
  
\section{\label{sect:mba}Applied to multi-bend achromat}

\begin{figure}[!ht]  
    \centering \includegraphics[width=.8\columnwidth]{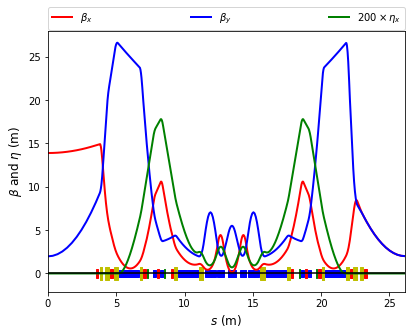}
    \caption{\label{fig:mbacell} Linear optics and magnet layout for
      one cell of an ESRF-EBS type hybrid 7BA lattice. The red blocks
      represent sextupoles. Six chromatic sextupoles grouped into five
      families inside two dispersive bumps are used to correct the
      chromaticity with three extra degrees of freedom. Four harmonic
      sextupoles and four dispersive octupoles (green blocks) are also
      available for nonlinear dynamics optimization.}
  \end{figure}

  Low-emittance light source ring design is now entering a new
  era. Various MBA-type lattices already reach diffraction-limited
  horizontal emittances to deliver much brighter X-ray beams. Like the DBA
  case, it is interesting to explore whether it is possible to design a
  nonlinear MBA lattice with two QIs. The ESRF-EBS type hybrid MBA
  lattice~\cite{farvacque2013} has been widely adopted by other
  facilities. It is also being considered as one of the options for future
  NSLS-II brightness upgrade. A preliminary 7-BA design is shown in
  Fig.~\ref{fig:mbacell}, which illustrates the linear optics for one
  cell.  Note that several reverse bends are
  incorporated~\cite{riemann2019, zhang2019}. The main parameters are
  listed in Table~\ref{tab:mba}.

  \begin{table}[!hbt]
   \centering
   \caption{Main parameters of the test hybrid MBA ring}
   \begin{tabular}{|p{0.4\columnwidth}|p{0.4\columnwidth}|}
     \hline
       \textbf{Parameters}       & \textbf{Values}   \\
       \hline
       \hline
       Hor. emit. ($pm$)         & 31                \\
       Natural chrom. (x/y)      & -125/-108         \\ 
       Tune (x/y)                & 73.19/28.62       \\ 
       Energy spread             & $7.1\times10^{-4}$ \\
       Damp. partition (x/y/s)   & 2.0/1.0/1.0       \\
     \hline
   \end{tabular}
   \label{tab:mba}
  \end{table}

\begin{figure}[!ht]
     \includegraphics[width=0.7\columnwidth]{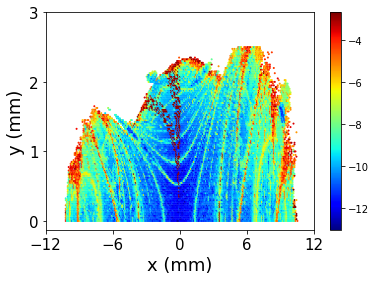}
     \caption{\label{fig:mbada} DA of the MBA lattice in the transverse
       $x-y$ plane colored with the tune diffusion.}
  \end{figure}

  A two-stage optimization has been implemented. First, the settings of
  the chromatic and harmonic sextupoles were optimized to correct the
  chromaticities, to minimize the fluctuations of the Courant-Snyder
  actions, and to maximize the ratio of the two leading frequency
  components. After this procedure, four octupoles inside the dispersive
  bumps were optimized to further minimize these objectives. The resulting
  DA is shown in Fig.~\ref{fig:mbada}.
  
\begin{figure}[!ht]
     \includegraphics[width=0.48\columnwidth]{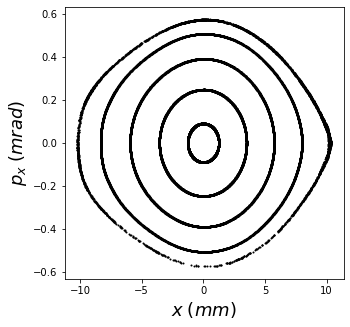}
     \includegraphics[width=0.48\columnwidth]{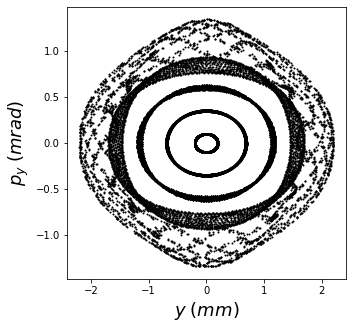}
     \caption{\label{fig:mbaphase} Simulated trajectories of the MBA
       lattice in the horizontal (left) and vertical (right) phase
       space. The vertical trajectories begin to smear out from thin tori
       gradually, but some patterns are visible.}
  \end{figure}

\begin{figure}[!ht]
     \includegraphics[width=0.7\columnwidth]{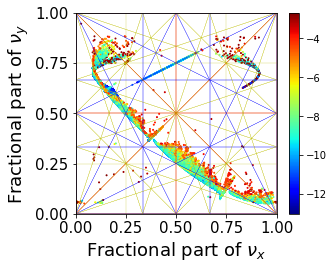}
     \caption{\label{fig:mbafma} Large amplitude-dependent tune-spread
       is observed in the MBA lattice constructed with QI.}
  \end{figure}

  The MBA lattice was found to be more challenging than the DBA lattice in
  constructing QI. The Courant-Snyder actions have larger fluctuations
  after optimization, especially in the vertical plane. Nevertheless,
  particle orbits are still confined to tori in the horizontal plane, as
  seen in Fig.~\ref{fig:mbaphase}. More importantly, a large
  amplitude-dependent tune-spread is observed within the tune footprint of
  the stable DA, as shown in Fig.~\ref{fig:mbafma}.
 
  At large amplitudes, the trajectories in the vertical phase space
  significantly deviate from the Courant-Snyder ellipse
  (Fig.~\ref{fig:mbaphase}).  However, this does not necessarily indicate
  that the tori are broken, as the 2D projections of 4D tori are observed.
  A more detailed analysis, for example, expanding the QI expression to
  high order polynomials for large-amplitude orbits in the vertical plane
  might be considered needed.

\section{\label{sect:period}Dependence on lattice location}

\begin{figure*}[!ht]
    \centering
    \includegraphics[width=0.9\textwidth]{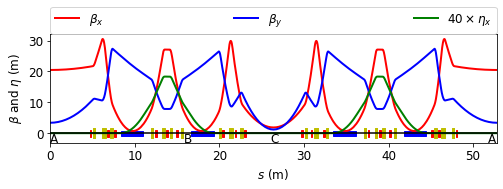}
    \includegraphics[width=0.21\textwidth]{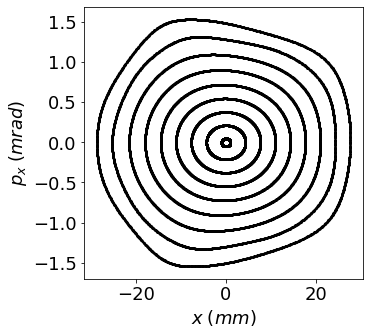}
    \includegraphics[width=0.20\textwidth]{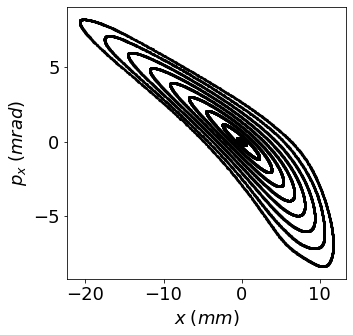}
    \includegraphics[width=0.21\textwidth]{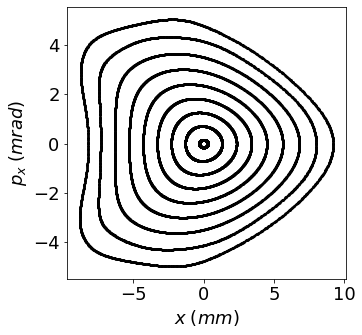}
    \includegraphics[width=0.21\textwidth]{ats0.png}
    \caption{\label{fig:anastigmat} Tori are near-circular only when
      observed at the long straight centers, but could be distorted at
      other locations. The bottom four subplots are the tori observed at
      the locations ``A-B-C-A'' respectively as marked in the top
      subplot.}
  \end{figure*}

  In the previous examples, the optimization procedure of
  Sect.~\ref{sect:invar} was performed using tracking data at a specific
  lattice location, e.g., at the center of the straight section.  Here,
  the orbits lie on invariant tori that are nearly circular when projected
  into the $\bar{x}-\bar{p}_x$ and $\bar{y}-\bar{p}_y$ planes. However, it
  is not surprising to observe that the invariant tori are distorted at
  other locations in the lattice.  Figure~\ref{fig:anastigmat} shows an
  example of distorted tori observed at a location inside an achromat of
  the NSLS-II DBA lattice. The one-cell map behaves like an ``optical
  anastigmat'', which has circular tori only at the periodic locations,
  i.e., its entrance and exit. Elsewhere in the lattice, distorted
  (perhaps broken) tori can be observed.
  
\section{\label{sect:decoh}decoherence due to nonlinearity}

\begin{figure}[!ht]
     \includegraphics[width=0.7\columnwidth]{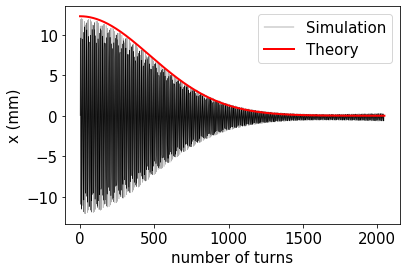}
     \caption{\label{fig:decoh} After being kicked, the centroid of the
       beam will show a decaying oscillation. The red line is known as the
       decoherence factor in Eq.~\eqref{eq:decoh}, and can be used to
       determine the local amplitude detuning coefficient $\mu$.}
  \end{figure}

  At the IOTA ring, its large amplitude-dependent tune-spread was found to
  provide a stabilization mechanism due to improved Landau
  damping~\cite{antipov2017_1}. Since constructed lattices with QI also
  provide a large nonlinear amplitude detuning, we used the DBA lattice to
  implement a 4-dimensional multi-particle simulation to illustrate this
  nonlinearity. A large tune spread due to the nonlinearity of betatron
  oscillations can be indirectly observed by kicking a bunched beam
  transversely~\cite{meller1987}.  A Gaussian distributed bunched beam was
  simulated by 2,000 macro-particles with the horizontal beam size
  $\sigma_x$.  After being kicked to an amplitude $Z=a\sigma_x$, the beam
  centroid will decay to the origin as the betatron phases of particles at
  different amplitudes decohere, which is described by the decoherence
  factor,
  \begin{equation}\label{eq:decoh}
    A(N)=\frac{1}{1+\theta^2}\exp
    \left[-\frac{Z^2}{2}\frac{\theta}{1+\theta^2}\right].
  \end{equation}
  Here, $N$ is the number of turns after being kicked, $\theta=4\pi\mu N$,
  and $\mu(Z)=\frac{d\nu}{da^2}$ is the local amplitude detuning
  coefficient at $Z$.  The phase space distribution of the beam filaments
  from a localized bunch to an annulus which occupies all betatron phases
  and the observed centroid of the beam will show a decaying oscillation
  as seen in Fig.~\ref{fig:decoh}. From the decoherence factor curve (the
  red line in Fig.~\ref{fig:decoh}), the local tune-shift-with-amplitude
  coefficient can be determined and confirmed to be consistent with the
  spectrum analysis of the single particle tracking data.
  
\section{\label{sect:summary}Discussion and summary}

\begin{figure}[!ht]
     \includegraphics[width=0.7\columnwidth]{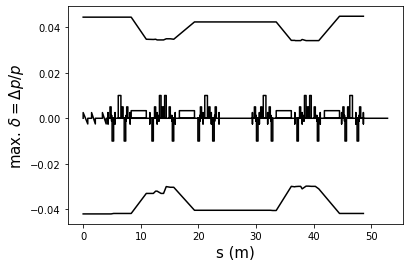}
     \includegraphics[width=0.7\columnwidth]{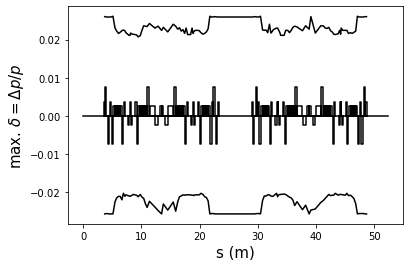}
     \caption{\label{fig:lma} Local momentum aperture of one supercell
       (including two mirror symmetric cells) for the constructed DBA
       (top) and MBA (bottom) lattices.}
  \end{figure}

  So far, only the on-momentum particle motion has been described. In
  practice, a sufficient off-momentum acceptance is required as well. In
  the two examples of Sect. \ref{sect:dba}-\ref{sect:mba}, on-momentum
  lattices were constructed with two quasi-invariants, and then the
  off-momentum acceptances were checked with tracking simulations. Both
  lattices were found to have sufficient local momentum aperture for the
  Touschek lifetime as shown in Fig.~\ref{fig:lma}. In case the momentum
  acceptance also needs to be optimized, the method can be expanded to
  include a set of off-momentum particles. For an off-momentum particle,
  the dispersive reference orbit needs to be computed first, and then the
  Courant-Snyder actions $J_{x,y}$ in (\ref{eq:quadraticJ}) need to be
  computed using the momentum-dependent Twiss parameters
  $\alpha(\delta,K_i),\;\beta(\delta,K_i)\;\text{and}\;\gamma(\delta,K_i)$. Note
  that the momentum-dependent Twiss parameters depend on the nonlinear
  magnets' excitation values $K_i$. For a given nonlinear magnet setting,
  the reference closed orbit and the corresponding momentum-dependent
  Twiss parameters need to be updated prior to the normalization.

  In this paper, we demonstrated that a conventional DBA or MBA lattice
  can be re-tuned to possess two approximate invariants by optimizing the
  settings of only the sextupoles and octupoles.  While the resulting DA
  is large (but finite), most particle trajectories are regular and
  confined to tori, and the amplitude-dependent betatron tunes are
  well-defined and stable. Like the lattice of the IOTA ring, a large
  nonlinear tune-spread exists that can provide enhanced Landau
  damping. The stop-band widths are observed narrow while the tune crosses
  low order resonance lines.
 
\begin{acknowledgments}
  We would like to thank L-H. Yu, B. Bacha, X. Yang and other NSLS-II
  colleagues for the stimulating discussion and/or productive experimental
  studies, M. Borland, R. Lindberg, Y-P. Sun, W. Cheng (ANL) and Y. Hao
  (MSU), E. Stern, A. Valishev, A. Romanov (FNAL), N. Kuklev (UChicago),
  G. Xu (IHEP) for the stimulating and collaborative discussions.  This
  research is supported by the U.S. Department of Energy (DoE) under
  Contract No. DE-SC0012704 (BNL) and DE-AC02-05CH11231 (LBNL). Co-authors
  from LBNL would like to acknowledge the support from the U.S. DoE Early
  Career Research Program under the Office of High Energy Physics.
\end{acknowledgments}
 
\section*{\label{sect:modulation}Appendix: Alternative optimization schemes}

\begin{figure}[!ht]
     \includegraphics[width=0.7\columnwidth]{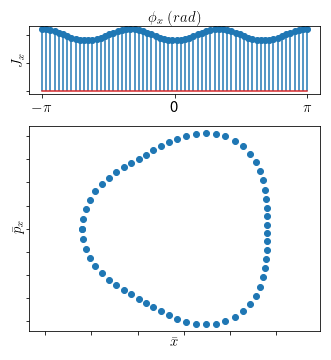}
     \caption{\label{fig:dbamodulation} Adding a periodic phase dependence
       (top) to the action $J_x$ can reshape a circular torus into a
       triangle-shaped one (bottom).}
  \end{figure}

  We can add additional terms to the Courant-Snyder actions $J_{x,y}$ in
  (\ref{eq:quadraticJ}), and use these in the optimization procedure
  described in Sect. \ref{sect:invar}, in order to construct a lattice
  with more complicated quasi-invariants. Here, we demonstrate
  this method by modifying the quasi-invariant $J_x$ to obtain
  triangle-shaped tori,
  \begin{align}{\label{eq:mod}}
    J_x(\phi_x)&=J_0+\Delta J_x(\phi_x)\nonumber\\ &=J_0\left\{1+\delta
    \sin\left[n\left(\phi_x-\frac{\pi}{2n}\right)\right]\right\},
  \end{align}  
  where $\delta=\frac{\Delta J_x}{J_x}$ and $n=3$. The phase dependence
  (top) of $J_x$ and the expected torus in the horizontal phase space
  (bottom) are illustrated in Fig.~\ref{fig:dbamodulation}.

\begin{figure}[!ht]
     \includegraphics[width=0.7\columnwidth]{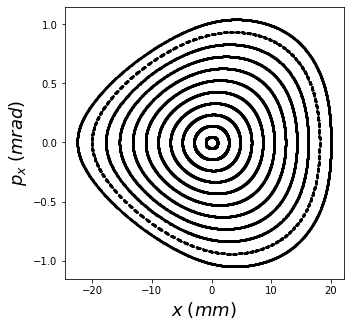}
     \caption{\label{fig:dbaxxptriangle} Particle trajectories are
       confined to triangle-shaped tori, as confirmed by the tracking
       simulation. A non-symmetric DA can be obtained with this
       technique.}
  \end{figure}

  After modifying the optimization procedure to minimize the fluctuations
  of (\ref{eq:mod}) in the DBA lattice, a new set of sextupole settings
  were obtained.  Tracking simulation confirmed that the tori have been
  reshaped as expected (Fig.~\ref{fig:dbaxxptriangle}). Such a phase space
  manipulation technique might be useful when a non-symmetric DA is
  desired. For example, a slightly larger inboard DA might be preferred to
  capture off-axis injected beams from that region.
  
  However, the phase dependence of (\ref{eq:mod}) can introduce additional
  frequency components into the betatron spectrum. For example, a weak
  component sitting near $1/3$ is visible through the NAFF analysis. If
  the betatron phase dependence is weak, it does not appear to spoil the
  overall performance of the nonlinear lattice. Note that the
  triangle-shaped tori are not caused by the betatron tune's approaching a
  third-order resonance, because the main betatron tune is around 0.22,
  and still far away from $1/3$.

\bibliography{quasi_invar.bib}

\begin{thebibliography}{29}%
\makeatletter
\providecommand \@ifxundefined [1]{%
 \@ifx{#1\undefined}
}%
\providecommand \@ifnum [1]{%
 \ifnum #1\expandafter \@firstoftwo
 \else \expandafter \@secondoftwo
 \fi
}%
\providecommand \@ifx [1]{%
 \ifx #1\expandafter \@firstoftwo
 \else \expandafter \@secondoftwo
 \fi
}%
\providecommand \natexlab [1]{#1}%
\providecommand \enquote  [1]{``#1''}%
\providecommand \bibnamefont  [1]{#1}%
\providecommand \bibfnamefont [1]{#1}%
\providecommand \citenamefont [1]{#1}%
\providecommand \href@noop [0]{\@secondoftwo}%
\providecommand \href [0]{\begingroup \@sanitize@url \@href}%
\providecommand \@href[1]{\@@startlink{#1}\@@href}%
\providecommand \@@href[1]{\endgroup#1\@@endlink}%
\providecommand \@sanitize@url [0]{\catcode `\\12\catcode `\$12\catcode
  `\&12\catcode `\#12\catcode `\^12\catcode `\_12\catcode `\%12\relax}%
\providecommand \@@startlink[1]{}%
\providecommand \@@endlink[0]{}%
\providecommand \url  [0]{\begingroup\@sanitize@url \@url }%
\providecommand \@url [1]{\endgroup\@href {#1}{\urlprefix }}%
\providecommand \urlprefix  [0]{URL }%
\providecommand \Eprint [0]{\href }%
\providecommand \doibase [0]{http://dx.doi.org/}%
\providecommand \selectlanguage [0]{\@gobble}%
\providecommand \bibinfo  [0]{\@secondoftwo}%
\providecommand \bibfield  [0]{\@secondoftwo}%
\providecommand \translation [1]{[#1]}%
\providecommand \BibitemOpen [0]{}%
\providecommand \bibitemStop [0]{}%
\providecommand \bibitemNoStop [0]{.\EOS\space}%
\providecommand \EOS [0]{\spacefactor3000\relax}%
\providecommand \BibitemShut  [1]{\csname bibitem#1\endcsname}%
\let\auto@bib@innerbib\@empty
\bibitem [{\citenamefont {Antipov}\ \emph
  {et~al.}(2017{\natexlab{a}})\citenamefont {Antipov} \emph
  {et~al.}}]{antipov2017}%
  \BibitemOpen
  \bibfield  {author} {\bibinfo {author} {\bibfnamefont {S.}~\bibnamefont
  {Antipov}} \emph {et~al.},\ }\bibfield  {title} {\enquote {\bibinfo {title}
  {{IOTA} (integrable optics test accelerator): facility and experimental beam
  physics program},}\ }\href@noop {} {\bibfield  {journal} {\bibinfo  {journal}
  {Journal of Instrumentation}\ }\textbf {\bibinfo {volume} {12}},\ \bibinfo
  {pages} {T03002} (\bibinfo {year} {2017}{\natexlab{a}})}\BibitemShut
  {NoStop}%
\bibitem [{\citenamefont {Danilov}(2008)}]{danilov2008}%
  \BibitemOpen
  \bibfield  {author} {\bibinfo {author} {\bibfnamefont {V.}~\bibnamefont
  {Danilov}},\ }\bibfield  {title} {\enquote {\bibinfo {title} {Practical
  solutions for nonlinear accelerator lattice with stable nearly regular
  motion},}\ }\href@noop {} {\bibfield  {journal} {\bibinfo  {journal}
  {Physical Review Special Topics-Accelerators and Beams}\ }\textbf {\bibinfo
  {volume} {11}},\ \bibinfo {pages} {114001} (\bibinfo {year}
  {2008})}\BibitemShut {NoStop}%
\bibitem [{\citenamefont {Danilov}\ and\ \citenamefont
  {Nagaitsev}(2010)}]{danilov2010}%
  \BibitemOpen
  \bibfield  {author} {\bibinfo {author} {\bibfnamefont {V.}~\bibnamefont
  {Danilov}}\ and\ \bibinfo {author} {\bibfnamefont {S.}~\bibnamefont
  {Nagaitsev}},\ }\bibfield  {title} {\enquote {\bibinfo {title} {Nonlinear
  accelerator lattices with one and two analytic invariants},}\ }\href@noop {}
  {\bibfield  {journal} {\bibinfo  {journal} {Physical Review Special
  Topics-Accelerators and Beams}\ }\textbf {\bibinfo {volume} {13}},\ \bibinfo
  {pages} {084002} (\bibinfo {year} {2010})}\BibitemShut {NoStop}%
\bibitem [{\citenamefont {Ruisard}\ \emph {et~al.}(2019)\citenamefont {Ruisard}
  \emph {et~al.}}]{ruisard2019}%
  \BibitemOpen
  \bibfield  {author} {\bibinfo {author} {\bibfnamefont {K}~\bibnamefont
  {Ruisard}} \emph {et~al.},\ }\bibfield  {title} {\enquote {\bibinfo {title}
  {Single-invariant nonlinear optics for a small electron recirculator},}\
  }\href@noop {} {\bibfield  {journal} {\bibinfo  {journal} {Physical Review
  Accelerators and Beams}\ }\textbf {\bibinfo {volume} {22}},\ \bibinfo {pages}
  {041601} (\bibinfo {year} {2019})}\BibitemShut {NoStop}%
\bibitem [{\citenamefont {Dragt}(2011)}]{dragt2011}%
  \BibitemOpen
  \bibfield  {author} {\bibinfo {author} {\bibfnamefont {A.}~\bibnamefont
  {Dragt}},\ }\bibfield  {title} {\enquote {\bibinfo {title} {Lie methods for
  nonlinear dynamics with applications to accelerator physics},}\ }\href@noop
  {} {\bibfield  {journal} {\bibinfo  {journal} {unpublished}\ } (\bibinfo
  {year} {2011})}\BibitemShut {NoStop}%
\bibitem [{\citenamefont {Chao}(1993)}]{chao1993}%
  \BibitemOpen
  \bibfield  {author} {\bibinfo {author} {\bibfnamefont {A.}~\bibnamefont
  {Chao}},\ }\href@noop {} {\emph {\bibinfo {title} {Physics of collective beam
  instabilities in high energy accelerators}}}\ (\bibinfo  {publisher}
  {Wiley},\ \bibinfo {year} {1993})\BibitemShut {NoStop}%
\bibitem [{\citenamefont {Stern}(2018)}]{stern2018}%
  \BibitemOpen
  \bibfield  {author} {\bibinfo {author} {\bibfnamefont {E.}~\bibnamefont
  {Stern}},\ }\href@noop {} {\emph {\bibinfo {title} {Suppression of
  Instabilities Generated by an Anti-Damper With a Nonlinear Magnetic Element
  in IOTA}}},\ \bibinfo {type} {Tech. Rep.}\ (\bibinfo  {institution} {Fermi
  National Accelerator Lab.},\ \bibinfo {year} {2018})\BibitemShut {NoStop}%
\bibitem [{\citenamefont {Yu}(2017)}]{yu2017}%
  \BibitemOpen
  \bibfield  {author} {\bibinfo {author} {\bibfnamefont {L.}~\bibnamefont
  {Yu}},\ }\bibfield  {title} {\enquote {\bibinfo {title} {Analysis of
  nonlinear dynamics by square matrix method},}\ }\href@noop {} {\bibfield
  {journal} {\bibinfo  {journal} {Physical Review Accelerators and Beams}\
  }\textbf {\bibinfo {volume} {20}},\ \bibinfo {pages} {034001} (\bibinfo
  {year} {2017})}\BibitemShut {NoStop}%
\bibitem [{\citenamefont {Borland}()}]{borland2017}%
  \BibitemOpen
  \bibfield  {author} {\bibinfo {author} {\bibfnamefont {M.}~\bibnamefont
  {Borland}},\ }\href@noop {} {}\bibinfo {howpublished} {personal
  communication}\BibitemShut {NoStop}%
\bibitem [{\citenamefont {Sun}\ and\ \citenamefont {Borland}(2017)}]{sun2017}%
  \BibitemOpen
  \bibfield  {author} {\bibinfo {author} {\bibfnamefont {Y.}~\bibnamefont
  {Sun}}\ and\ \bibinfo {author} {\bibfnamefont {M.}~\bibnamefont {Borland}},\
  }\bibfield  {title} {\enquote {\bibinfo {title} {{Comparison of Nonlinear
  Dynamics Optimization Methods for APS-U}},}\ }in\ \href {\doibase
  10.18429/JACoW-NAPAC2016-WEPOB15} {\emph {\bibinfo {booktitle} {{2nd North
  American Particle Accelerator Conference}}}}\ (\bibinfo {year} {2017})\ p.\
  \bibinfo {pages} {WEPOB15}\BibitemShut {NoStop}%
\bibitem [{\citenamefont {Liouville}(1855)}]{liouville1855}%
  \BibitemOpen
  \bibfield  {author} {\bibinfo {author} {\bibfnamefont {J.}~\bibnamefont
  {Liouville}},\ }\bibfield  {title} {\enquote {\bibinfo {title} {Note sur
  l'int{\'e}gration des {\'e}quations diff{\'e}rentielles de la dynamique,
  pr{\'e}sent{\'e}e au bureau des longitudes le 29 juin 1853.}}\ }\href@noop {}
  {\bibfield  {journal} {\bibinfo  {journal} {Journal de Math{\'e}matiques
  pures et appliqu{\'e}es}\ ,\ \bibinfo {pages} {137--138}} (\bibinfo {year}
  {1855})}\BibitemShut {NoStop}%
\bibitem [{\citenamefont {Arnold}()}]{arnold1989}%
  \BibitemOpen
  \bibfield  {author} {\bibinfo {author} {\bibfnamefont {V.}~\bibnamefont
  {Arnold}},\ }\bibfield  {title} {\enquote {\bibinfo {title} {{Mathematical
  Methods of Classical Mechanics, Springer--Verlag, Berlin and New York,
  1989}},}\ }\href@noop {} {\ }\BibitemShut {NoStop}%
\bibitem [{\citenamefont {Courant}\ and\ \citenamefont
  {Snyder}(1958)}]{courant1958}%
  \BibitemOpen
  \bibfield  {author} {\bibinfo {author} {\bibfnamefont {E.}~\bibnamefont
  {Courant}}\ and\ \bibinfo {author} {\bibfnamefont {H.}~\bibnamefont
  {Snyder}},\ }\bibfield  {title} {\enquote {\bibinfo {title} {Theory of the
  alternating-gradient synchrotron},}\ }\href {\doibase
  https://doi.org/10.1016/0003-4916(58)90012-5} {\bibfield  {journal} {\bibinfo
   {journal} {Annals of Physics}\ }\textbf {\bibinfo {volume} {3}},\ \bibinfo
  {pages} {1 -- 48} (\bibinfo {year} {1958})}\BibitemShut {NoStop}%
\bibitem [{\citenamefont {Yoshida}(1990)}]{yoshida1990}%
  \BibitemOpen
  \bibfield  {author} {\bibinfo {author} {\bibfnamefont {H.}~\bibnamefont
  {Yoshida}},\ }\bibfield  {title} {\enquote {\bibinfo {title} {{Construction
  of higher order symplectic integrators}},}\ }\href {\doibase
  10.1016/0375-9601(90)90092-3} {\bibfield  {journal} {\bibinfo  {journal}
  {Phys. Lett.}\ }\textbf {\bibinfo {volume} {A150}},\ \bibinfo {pages}
  {262--268} (\bibinfo {year} {1990})}\BibitemShut {NoStop}%
\bibitem [{\citenamefont {Chao}(2002)}]{chao2002}%
  \BibitemOpen
  \bibfield  {author} {\bibinfo {author} {\bibfnamefont {A.}~\bibnamefont
  {Chao}},\ }\href@noop {} {\emph {\bibinfo {title} {{Lecture Notes on Topics
  in Accelerator Physics}}}}\ (\bibinfo  {publisher} {SLAC},\ \bibinfo
  {address} {Stanford, CA},\ \bibinfo {year} {2002})\BibitemShut {NoStop}%
\bibitem [{\citenamefont {Laskar}(2003)}]{laskar2003}%
  \BibitemOpen
  \bibfield  {author} {\bibinfo {author} {\bibfnamefont {J.}~\bibnamefont
  {Laskar}},\ }\bibfield  {title} {\enquote {\bibinfo {title} {{Frequency Map
  Analysis and Particle Accelerators}},}\ }in\ \href@noop {} {\emph {\bibinfo
  {booktitle} {20th Particle Accelerator Conference (PAC 03)}}}\ (\bibinfo
  {year} {2003})\ p.\ \bibinfo {pages} {378}\BibitemShut {NoStop}%
\bibitem [{\citenamefont {Deb}(2001)}]{deb2001}%
  \BibitemOpen
  \bibfield  {author} {\bibinfo {author} {\bibfnamefont {K.}~\bibnamefont
  {Deb}},\ }\href@noop {} {\emph {\bibinfo {title} {Multi-Objective
  Optimization Using Evolutionary Algorithms}}}\ (\bibinfo  {publisher}
  {Wiley},\ \bibinfo {year} {2001})\BibitemShut {NoStop}%
\bibitem [{\citenamefont {Edwards}\ and\ \citenamefont
  {Teng}(1973)}]{edwards1973}%
  \BibitemOpen
  \bibfield  {author} {\bibinfo {author} {\bibfnamefont {D.}~\bibnamefont
  {Edwards}}\ and\ \bibinfo {author} {\bibfnamefont {L.}~\bibnamefont {Teng}},\
  }\bibfield  {title} {\enquote {\bibinfo {title} {Parametrization of linear
  coupled motion in periodic systems},}\ }\href@noop {} {\bibfield  {journal}
  {\bibinfo  {journal} {IEEE Transactions on nuclear science}\ }\textbf
  {\bibinfo {volume} {20}},\ \bibinfo {pages} {885--888} (\bibinfo {year}
  {1973})}\BibitemShut {NoStop}%
\bibitem [{\citenamefont {{BNL}}()}]{bnl2013}%
  \BibitemOpen
  \bibfield  {author} {\bibinfo {author} {\bibnamefont {{BNL}}},\ }\href@noop
  {} {}\bibinfo {howpublished}
  {\url{https://www.bnl.gov/nsls2/project/PDR/}}\BibitemShut {NoStop}%
\bibitem [{\citenamefont {Borland}(2000)}]{borland2000}%
  \BibitemOpen
  \bibfield  {author} {\bibinfo {author} {\bibfnamefont {M.}~\bibnamefont
  {Borland}},\ }\bibfield  {title} {\enquote {\bibinfo {title} {{elegant: A
  Flexible SDDS-Compliant Code for Accelerator Simulation}},}\ }in\ \href@noop
  {} {\emph {\bibinfo {booktitle} {{Advanced Photon Source LS-287}}}}\
  (\bibinfo {year} {2000})\BibitemShut {NoStop}%
\bibitem [{\citenamefont {Schoch}(1957)}]{schoch1957}%
  \BibitemOpen
  \bibfield  {author} {\bibinfo {author} {\bibfnamefont {A.}~\bibnamefont
  {Schoch}},\ }\bibfield  {title} {\enquote {\bibinfo {title} {Theory of
  non-linear perturbations of betatron oscillations},}\ }\href@noop {}
  {\bibfield  {journal} {\bibinfo  {journal} {CERN Report}\ ,\ \bibinfo {pages}
  {57--21}} (\bibinfo {year} {1957})}\BibitemShut {NoStop}%
\bibitem [{\citenamefont {Wiedemann}(2015)}]{wiedemann2015}%
  \BibitemOpen
  \bibfield  {author} {\bibinfo {author} {\bibfnamefont {H.}~\bibnamefont
  {Wiedemann}},\ }\href@noop {} {\emph {\bibinfo {title} {Resonances. In:
  Particle accelerator physics}}}\ (\bibinfo  {publisher} {Springer Nature},\
  \bibinfo {year} {2015})\BibitemShut {NoStop}%
\bibitem [{\citenamefont {Cheng}\ \emph {et~al.}(2018)\citenamefont {Cheng},
  \citenamefont {Bacha}, \citenamefont {Li},\ and\ \citenamefont
  {Teytelman}}]{cheng2018}%
  \BibitemOpen
  \bibfield  {author} {\bibinfo {author} {\bibfnamefont {W.}~\bibnamefont
  {Cheng}}, \bibinfo {author} {\bibfnamefont {B.}~\bibnamefont {Bacha}},
  \bibinfo {author} {\bibfnamefont {Y.}~\bibnamefont {Li}}, \ and\ \bibinfo
  {author} {\bibfnamefont {D.}~\bibnamefont {Teytelman}},\ }\bibfield  {title}
  {\enquote {\bibinfo {title} {{Developments of Bunch by Bunch Feedback System
  at NSLS-II Storage Ring}},}\ }in\ \href {\doibase
  10.18429/JACoW-IPAC2018-WEPAF011} {\emph {\bibinfo {booktitle} {{9th
  International Particle Accelerator Conference}}}}\ (\bibinfo {year}
  {2018})\BibitemShut {NoStop}%
\bibitem [{\citenamefont {Cheng}\ \emph {et~al.}(2016)\citenamefont {Cheng}
  \emph {et~al.}}]{cheng2016}%
  \BibitemOpen
  \bibfield  {author} {\bibinfo {author} {\bibfnamefont {W.}~\bibnamefont
  {Cheng}} \emph {et~al.},\ }\bibfield  {title} {\enquote {\bibinfo {title}
  {Experimental study of single bunch instabilities at nsls-ii storage ring},}\
  }\href@noop {} {\bibfield  {journal} {\bibinfo  {journal} {Proc. IPAC’16}\
  } (\bibinfo {year} {2016})}\BibitemShut {NoStop}%
\bibitem [{\citenamefont {Farvacque}\ \emph {et~al.}(2013)\citenamefont
  {Farvacque} \emph {et~al.}}]{farvacque2013}%
  \BibitemOpen
  \bibfield  {author} {\bibinfo {author} {\bibfnamefont {L.}~\bibnamefont
  {Farvacque}} \emph {et~al.},\ }\bibfield  {title} {\enquote {\bibinfo {title}
  {A low-emittance lattice for the {ESRF}},}\ }\href@noop {} {\bibfield
  {journal} {\bibinfo  {journal} {Proc. IPAC’13}\ ,\ \bibinfo {pages}
  {79--81}} (\bibinfo {year} {2013})}\BibitemShut {NoStop}%
\bibitem [{\citenamefont {Riemann}\ and\ \citenamefont
  {Streun}(2019)}]{riemann2019}%
  \BibitemOpen
  \bibfield  {author} {\bibinfo {author} {\bibfnamefont {B.}~\bibnamefont
  {Riemann}}\ and\ \bibinfo {author} {\bibfnamefont {A.}~\bibnamefont
  {Streun}},\ }\bibfield  {title} {\enquote {\bibinfo {title} {Low emittance
  lattice design from first principles: Reverse bending and longitudinal
  gradient bends},}\ }\href@noop {} {\bibfield  {journal} {\bibinfo  {journal}
  {Physical Review Accelerators and Beams}\ }\textbf {\bibinfo {volume} {22}},\
  \bibinfo {pages} {021601} (\bibinfo {year} {2019})}\BibitemShut {NoStop}%
\bibitem [{\citenamefont {Zhang}\ and\ \citenamefont
  {Huang}(2019)}]{zhang2019}%
  \BibitemOpen
  \bibfield  {author} {\bibinfo {author} {\bibfnamefont {T.}~\bibnamefont
  {Zhang}}\ and\ \bibinfo {author} {\bibfnamefont {X.}~\bibnamefont {Huang}},\
  }\bibfield  {title} {\enquote {\bibinfo {title} {Numerical optimization of a
  low emittance lattice cell},}\ }\href@noop {} {\bibfield  {journal} {\bibinfo
   {journal} {Nuclear Instruments and Methods in Physics Research Section A:
  Accelerators, Spectrometers, Detectors and Associated Equipment}\ }\textbf
  {\bibinfo {volume} {923}},\ \bibinfo {pages} {55--63} (\bibinfo {year}
  {2019})}\BibitemShut {NoStop}%
\bibitem [{\citenamefont {Antipov}\ \emph
  {et~al.}(2017{\natexlab{b}})\citenamefont {Antipov}, \citenamefont
  {Nagaitsev} \emph {et~al.}}]{antipov2017_1}%
  \BibitemOpen
  \bibfield  {author} {\bibinfo {author} {\bibfnamefont {S.}~\bibnamefont
  {Antipov}}, \bibinfo {author} {\bibfnamefont {S.}~\bibnamefont {Nagaitsev}},
  \emph {et~al.},\ }\bibfield  {title} {\enquote {\bibinfo {title}
  {Single-particle dynamics in a nonlinear accelerator lattice: attaining a
  large tune spread with octupoles in iota},}\ }\href@noop {} {\bibfield
  {journal} {\bibinfo  {journal} {Journal of Instrumentation}\ }\textbf
  {\bibinfo {volume} {12}},\ \bibinfo {pages} {P04008} (\bibinfo {year}
  {2017}{\natexlab{b}})}\BibitemShut {NoStop}%
\bibitem [{\citenamefont {Meller}\ \emph {et~al.}(1987)\citenamefont {Meller},
  \citenamefont {Chao} \emph {et~al.}}]{meller1987}%
  \BibitemOpen
  \bibfield  {author} {\bibinfo {author} {\bibfnamefont {R.}~\bibnamefont
  {Meller}}, \bibinfo {author} {\bibfnamefont {A.}~\bibnamefont {Chao}},  \emph
  {et~al.},\ }\bibfield  {title} {\enquote {\bibinfo {title} {{Decoherence of
  Kicked Beams}},}\ }\href@noop {} {\bibfield  {journal} {\bibinfo  {journal}
  {SSC-N-360}\ } (\bibinfo {year} {1987})}\BibitemShut {NoStop}%
\end{thebibliography}%

\end{document}